\documentclass[aps,pre,twocolumn,float]{revtex4}
\usepackage{amsmath,bm,epsfig}

\def\Fbox#1{\vskip1ex\hbox to 8.5cm{\hfil\fboxsep0.3cm\fbox{%
  \parbox{8.0cm}{#1}}\hfil}\vskip1ex\noindent}  

\newcommand{\B}[1]{{\bm{#1}}}
\let \= \equiv \let\*\cdot \let\~\widetilde \let\^\widehat \let\-\overline

\begin{document}
\title{Size of Plastic Events in Strained Amorphous Solids at Finite Temperatures}
\author{H.G.E. Hentschel$^*$, Smarajit Karmakar, Edan Lerner and Itamar Procaccia}
\affiliation{Department of Chemical Physics, The Weizmann
Institute of Science, Rehovot 76100, Israel \\
$^*$ Dept. of Physics, Emory University, Atlanta Ga. 30322 }
\date{\today}
\begin{abstract}
We address the system-size dependence of typical plastic flow events when an amorphous solid is put under a fixed external strain rate at a finite temperature. For system sizes that are accessible to numerical simulations at reasonable strain rates and at low temperatures the magnitude of plastic events grows with the system size. We explain however that this must be a finite size effect; for larger systems there exist two cross-over length-scales $\xi_1$ and $\xi_2$, the first determined by the elastic time-scale and the second by the thermal energy-scale. For system of linear size $L$ larger than any of these scales the magnitude of plastic events must be bounded. For systems of size $L\gg \xi$ there must exist $(L/\xi)^d$ uncorrelated plastic events which occur simultaneously. We present a scaling theory that culminates with the dependence of the cross-over scales on temperature and strain rate. Finally we relate these findings to the temperature and size dependence of the stress fluctuations. We comment on the importance of these considerations for theories of elasto-plasticity.
\end{abstract}
\maketitle

{\bf Introduction:} The issues of the statistical correlations of plastic flow events in strained amorphous solids are central to the possible form of the dynamical theory of elasto-plasticity \cite{79AK,79Arg,82AS,98FL,98Sol,07BLP}. As such they were at the center of extensive research in recent years \cite{04ML,06TLB,06ML,09MR,09LP}. The crucial question is whether these events are spatially localized and statistically independent, as assumed often in the theoretical development, or are they statistically correlated to form extended events that depend on the system size.
Of particular relevance to the present Letter is Ref. \cite{09LC} in which the authors studied the question for zero temperature as a function of the strain rate. At low strain rates $\dot \gamma$ the plastic events were shown to be spatially correlated with a system size dependence. At high strain rates (compared to elastic relaxation times) the correlation were cut-off proportional to $\dot\gamma^{-1/d}$ where
$d$ is the space dimension. Two crucial questions that remain are (i) what is the effect of temperature on this issue. Should temperature fluctuations also cut-off the statistical correlations? and (ii) if temperature effects do cut off the magnitude of plastic flow events, which of the cut-offs dominates at a given temperature and strain rate?

The aim of this Letter is to address these two questions. We will show that temperature effects are as important, if not
more important, in checking the magnitude of plastic events as the effect of a finite $\dot\gamma$. We will present below
some quantitative estimates of the various effects to compare their efficacy in bounding the magnitude of
plastic flow events at a given temperature and strain rate.

{\bf Summary of the Athermal, Quasi-static Simulations :} At athermal conditions $T=0$ an amorphous solid subjected to very slow strain rate (quasi-static in the limit) tends to set up an elasto-plastic steady state in which short elastic intervals in which the energy and the stress slowly increase are interrupted by plastic flow events during which the energy and the stress decrease on the short time scale of elastic relaxation. During the steady state one can measure accurately the average stress drops $\langle \Delta \sigma \rangle$ or the average energy drops $\langle \Delta U \rangle$. In both two-dimensions \cite{09LP}
and three-dimensions \cite{07BSLJ} it was found that these averages depend on the total number of particles as power-laws,
\begin{equation}
\langle \Delta U \rangle = \bar \epsilon N^\alpha\ , \quad \langle \Delta \sigma \rangle = s N^\beta \ , \label{scaling}
\end{equation}
with $\alpha>0$ and $\beta<0$, where $\bar\epsilon$ is the mean energy drop per particle, and $s$ is a stress scale to be computed below. A scaling relation $\alpha-\beta=1$ follows from the average energy balance equation, cf. \cite{09LP}
\begin{equation}
\frac{\sigma_Y \langle \Delta \sigma \rangle}{\mu} V =  \langle \Delta U \rangle \ , \label{bal}
\end{equation}
where $\sigma_Y$ is the flow stress (the mean stress in the athermal steady state) and $\mu$ is the shear modulus. The actual values of the exponents $\alpha$ and $\beta$ can depend on the details of the inter-particle potential. Typical values of $\alpha$ are a bit less than 0.4 in two dimensions \cite{09LP} and a bit more than 0.4 in three dimensions \cite{07BSLJ}. In Ref. \cite{09LP} it was shown that the number of particles participating in a plastic flow events scales like $\langle \Delta U \rangle$.

{\bf The effect of finite strain rate:} As said in the introduction, Ref. \cite{09LC} showed that finite strain rates may cut-off the magnitude of plastic flow events. To understand this effect we start by substituting Eq. (\ref{scaling}) in Eq. (\ref{bal}) to obtain the scale $s$,
\begin{equation}
s= \frac{\bar \epsilon\mu}{\sigma_Y \langle \lambda \rangle ^d} =\frac{\bar\epsilon\mu\rho}{\sigma_Y m} \ .
\end{equation}
Consider next the rate at which work is being done at the system and balance it by the energy dissipation in the steady state,
\begin{equation}
\sigma_Y \dot \gamma V =\langle \Delta U \rangle/\tau_{\rm pl} \ , \label{work}
\end{equation}
where $\tau_{\rm pl}$ is the average time between plastic flow events. This time is estimated as the elastic rise time which is
\begin{equation}
\tau_{\rm pl} \sim \frac{\langle\Delta \sigma \rangle}{\mu\dot\gamma} \sim \frac{\bar\epsilon N^\beta}{\sigma_Y \langle \lambda \rangle^d \dot\gamma} \ .
\end{equation}
We increase our confidence in this estimate by substituting it into Eq. (\ref{work}) together with the other estimates, to find perfect consistency.

Next we note that $\tau_{\rm pl}$ decreases when $N$ increases. On the other hand there exists another crucial time scale in the system, which is the elastic relaxation time
\begin{equation}
\tau_{\rm el} \sim L/c
\end{equation}
 where $c$ is the speed of sound $c=\sqrt{\mu/\rho}$. Obviously this time scale {\em increases} with $N$ like $N^{1/d}$. There will be therefore a typical scale $\xi_1$ such that for a system of scale $L=\xi_1$ these times cross. At that size the system cannot equilibrate its elastic energy before another event is triggered, and multiple avalanches must be occurring simultaneously in different parts of the system, each of which has a bounded magnitude.
We estimate $\xi_1$ from $\tau_{\rm el} \sim \tau_{\rm pl}$, finding
\begin{equation}
(\xi_1/c) \sim \frac{\epsilon [N(\xi_1)]^\beta}{\sigma_Y \langle \lambda \rangle^d\dot\gamma}\sim \frac{\epsilon [\xi/\langle\lambda\rangle]^{d\beta}}{\sigma_Y \langle \lambda \rangle^d\dot\gamma} \ .
\end{equation}
Using now the obvious fact that $N(\xi_1) \sim (\xi_1/\langle \lambda \rangle)^d$ we compute
\begin{equation}
\frac{\xi_1}{\langle \lambda \rangle} \sim \left[\left(\frac{\bar\epsilon}{\sigma_Y \langle \lambda \rangle^d}\right)\, \left(\frac{c}{ \langle \lambda \rangle\dot\gamma}\right) \right]^{1/(1-\beta d)}
\end{equation}
We first observe the singularity for quasi-static strain when $\dot \gamma\to 0$, where $\xi_1$ tends to infinity, in agreement with the results of quasi-static calculations. Thus at low temperatures, before the thermal energy scale becomes important, the size of plastic flow events can be huge indeed. We show next that thermal effects put a much more stringent bounds on the magnitude of plastic flow events.

{\bf The effect of finite temperatures:} The typical scale of thermal fluctuations is $k_B T N$ where $k_B$ is Boltzmann's
constant. Comparing with the $N$ dependence of the typical energy drop due to plastic flow events, we see that the
former increases faster with $N$, and it will catch up when
\begin{equation}
\bar \epsilon N^\alpha \sim N k_B T  \ . \label{cross2}
\end{equation}
This equality will hold when the system size $L=\xi_2$, where $(\xi_2/\langle \lambda\rangle)^d = N$. Substituting the last
equality in Eq. (\ref{cross2}) and then solving for $\xi_2$ we find
\begin{equation}
\frac{\xi_2}{\langle \lambda \rangle} = \left[\frac{k_BT}{\bar\epsilon}\right]^{1/d\beta} \ .
\end{equation}
Recalling that $\beta = \alpha-1$ is negative, we again notice the singularity at $T\to 0$ in agreement with the
athermal quasi-static simulations.

\begin{figure}
\centering
\hskip -0.8 cm
\includegraphics[scale = 0.395]{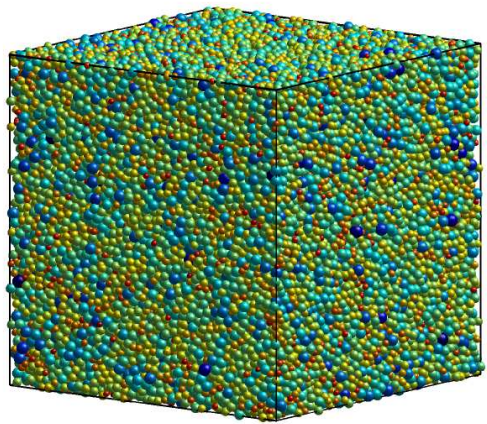}
\caption{A typical equilibrium configuration with 65,536 particles. The particles are all point objects, and the ball around each particle is of radius $\lambda_i$.}
\label{system}
\end{figure}

{\bf A Model Glass Example}: To put some size estimates on these crucial length-scales, and to test their consequences, we need to choose a
model glass. To this aim we employ a model system with point particles of equal mass $m$ and positions $\B r_i$ in three-dimensions, interacting via a pair-wise interaction potentials. In our three-dimensional simulations each particle $i$ is assigned an interaction parameter $\lambda_i$ from a normal distribution with mean $\langle \lambda \rangle$. The variance is governed by the poly-dispersity parameter $\Delta = 15\%$ where $\Delta^2 = \frac{\langle \left(\lambda_i -
\langle \lambda \rangle \right)^2\rangle}{\langle \lambda \rangle^2} $. With the definitions $r_{ij}=|\B r_i-\B r_j|$ and $\lambda_{ij} = \case{1}{2}(\lambda_i + \lambda_j)$, the potential assumes the form
\begin{widetext}
\begin{equation}
U(r_{ij}) =
\left\{
\begin{array}{ccl}
\!\!\! \epsilon\left[\left(\frac{\lambda_{ij}}{r_{ij}}\right)^{k}\!\! -\!\!\frac{k(k+2)}{8}
\left( \frac{B_0}{k} \right)^{\frac{k+4}{k+2}}\left(\frac{r_{ij}}{\lambda_{ij}}\right)^4
+ \frac{B_0(k+4)}{4}\left(\frac{r_{ij}}{\lambda_{ij}}\right)^2
-\frac{(k+2)(k+4)}{8}\left( \frac{B_0}{n} \right)^{\frac{k}{k+2}}\right] & , & r_{ij} \le
\lambda_{ij}\left( \frac{k}{B_0} \right)^{\frac{1}{k+2}} \\
0 & , & r_{ij} >
\lambda_{ij}\left( \frac{k}{B_0} \right)^{\frac{1}{k+2}}
\label{potential}
\end{array}
\right\}\ ,
\end{equation}
\end{widetext}
In our two dimensional simulations below we use the same potential but choose a binary mixture model with a `large' and
a `small' particles such that $\lambda_{LL} =1.4$, $\lambda_{LS}=1.18$ and $\lambda_{SS}=1.00$.
Below the units of length, energy, mass and temperature are $\langle \lambda\rangle$, $\epsilon$, $m$ and $\epsilon/k_B$ where $k_B$ is Boltzmann's constant. The time units $\tau_0$ are accordingly $\tau_0=\sqrt{(m\langle \lambda\rangle^2/\epsilon})$. The motivation of this somewhat lengthy form of the potential is to have continuous first and second derivatives at the built-in cutoff of $ r_{ij} = \lambda_{ij}\left(k/B_0 \right)^{\frac{1}{k+2}}$. In the present simulations we chose $k=10$, $B_0=0.2$. The choice of a quartic rather than a quadratic correction term is motivated by numerical speed considerations, avoiding the calculation of square roots. In the 3D simulations below the mass density $\rho\equiv m N/V=1.3$, whereas in 2D $\rho=0.85$. In all cases the boundary conditions are periodic. In Fig. \ref{system} we present a typical 3D equilibrium configuration of the system with $N=65536$. We measured for this 3D system the shear modulus $\mu=15.7$ and therefore the speed of sound is $c\approx 3.5$. The value
of $\sigma_Y$ at $T=0$ is about 0.7 and the typical value of $\sigma_\infty$ at higher temperatures is of the order
of $0.5$.

The estimate of $\xi_1$ depends of course on $\dot\gamma$. In our 3D simulations we have used $\dot\gamma=5\times 10^{-5}$,
and for the given values of the speed of sound and of $\sigma_Y$ we estimate $\xi_1/\langle\lambda\rangle\sim 2\times 10^5$ which translates to about $10^{16}$ particles. Obviously this system size is hugely beyond the capabilities of molecular simulations. One could in principle increase $\dot \gamma$, but not beyond $\sigma_y/(\sqrt{\rho\mu} L)$ \cite{foot}. It therefore remains elusive to demonstrate the cross-over due to the elastic time-scale in numerical simulations. Nevertheless one should remember in
any attempt of developing an athermal theory of elasto-plasticity that the plastic flow events are very large, a fact that cannot be disregarded with impunity.

The cross-over scale due to thermal energies is very well within the range of system size available in numerical
simulations. Making the plausible estimate $\bar\epsilon \approx \epsilon$ we see that already at $T=10^{-3}$
$\xi_2/\langle\lambda\rangle$ is estimated (for $\beta =-8/15$ in 3 dimensions \cite{07BSLJ}) as $\xi_2\approx 10^2$, which
translates to just 1 million particles. For $T=10^{-2}$ this estimate drops down to about 1000 particles.
Thus we expect a very rapid cross-over from correlated avalanches to un-correlated ones as the temperature rises
above $10^{-3}$.

\begin{figure}
\centering
\hskip -1.0 cm
\includegraphics[scale = 0.45]{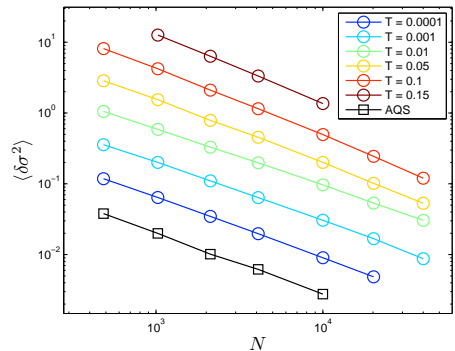}
\caption{The variance of the stress fluctuations as a function of the system size $N$ for a 2D system and for various temperatures. The first power-law (data in squares) is obtained under athermal quasi-static conditions where we determine for the present model $\beta=-0.61$, $\theta=-0.40$. The other plots go up in temperature as
indicated. The plots are displaced by a fixed amount for clarity. Note that the slope decreases (becoming more negative) as the temperature increases}
\label{deltasig1}
\end{figure}

\begin{figure}
\centering
\hskip -1.0 cm
\includegraphics[scale = 0.45]{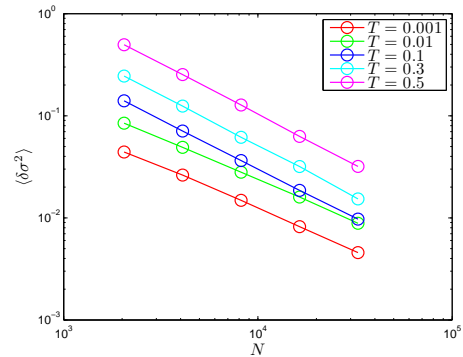}
\caption{The variance of the stress fluctuations as a function of the system size $N$ for a 3D system and for various temperatures. The plots are displaced by a fixed amount for clarity. Note that the slope decreases (becoming more negative) as the temperature increases}
\label{deltasig2}
\end{figure}

{\bf Demonstration of the Thermal Cross-over: }A very interesting and direct way of demonstrating the cross-over due to thermal effects is provided by measurements
of the variance of the stress fluctuations as a function of the temperature and the system size. This variance is
defined by
\begin{equation}
\langle \delta \sigma^2\rangle \equiv \langle (\sigma -\sigma_\infty)^2\rangle \ , \label{fluct}
\end{equation}
where $\sigma_\infty$ is the mean stress in the thermal steady state. In Fig. \ref{deltasig1} and \ref{deltasig2} we display 2D and 3D measurements of this quantity which is obtained by averaging the square of the microscopic stress fluctuations in long stretches of elasto-plastic steady-states of the models described above at a fixed $\dot\gamma=2.5\times 10^{-5}$
in 2D and $\dot\gamma=5\times 10^{-5}$ in 3D.
It is evident that the variance of the stress fluctuations decreases as a function of $N$. Under quasi-static and
athermal conditions the dependence is a power-law
\begin{equation}
\langle \delta \sigma^2\rangle \sim N^{2\theta} \ ,
\end{equation}
where $\theta\approx -0.4$ both in 2D and 3D. One should notice the difference between the exponent characterizing the $N$
dependence of $\sqrt{\langle \delta \sigma^2\rangle}$ and of the athermal mean plastic stress drop $\langle \Delta \sigma\rangle$, in the sense that $\theta\ne \beta$. This difference is due to very strong correlations between elastic increases and plastic drops. At higher temperatures the data in Figs. \ref{deltasig1} and \ref{deltasig2} indicate a clear cross-over to independent stress fluctuations in which
\begin{equation}
\langle \delta \sigma^2\rangle \sim N^{-1}\ , \quad {\rm for~high~temperatures} \ .
\end{equation}

\begin{figure}
\centering
\hskip -1.0 cm
\includegraphics[scale = 0.45]{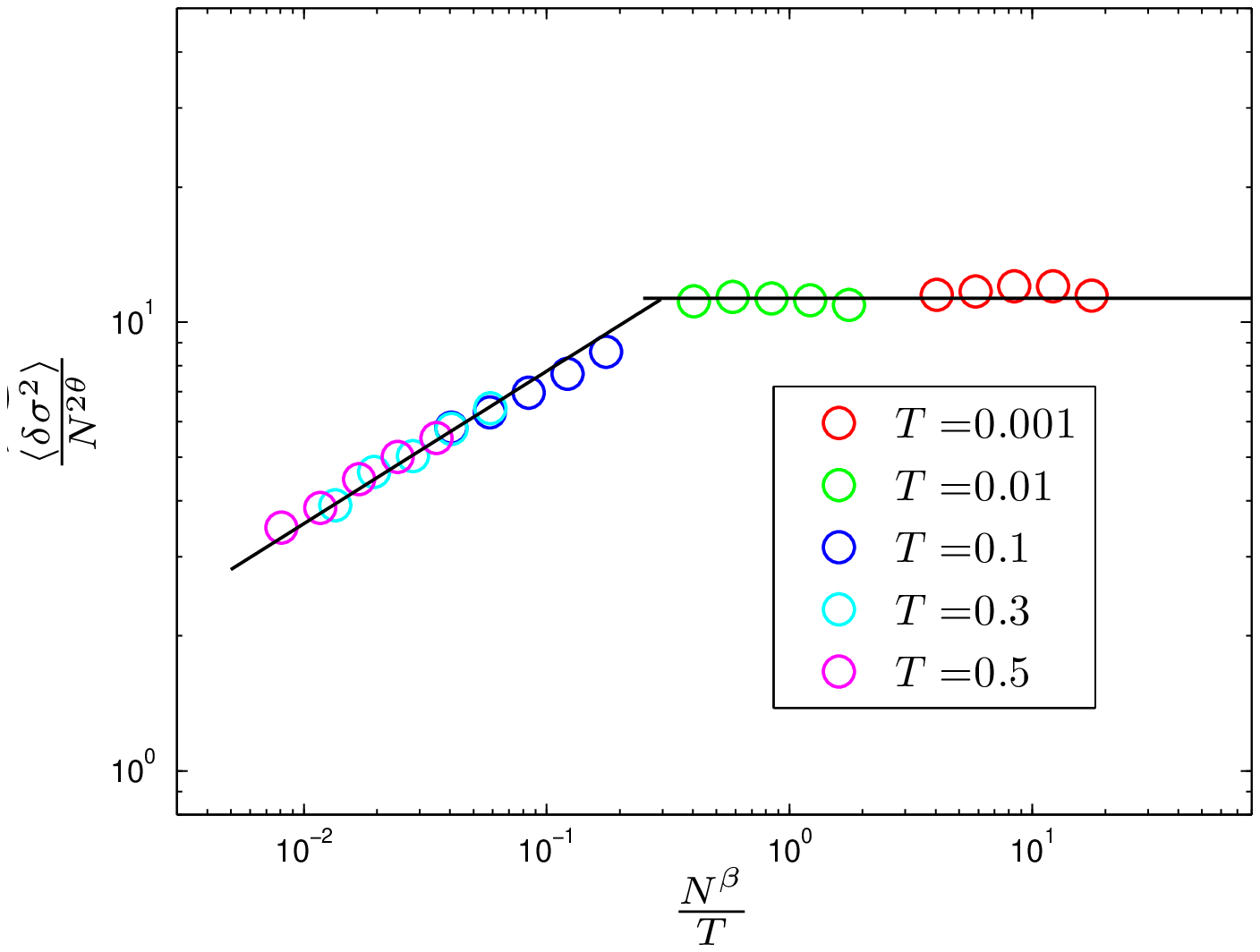}
\centering
\hskip -1.4 cm
\includegraphics[scale = 0.45]{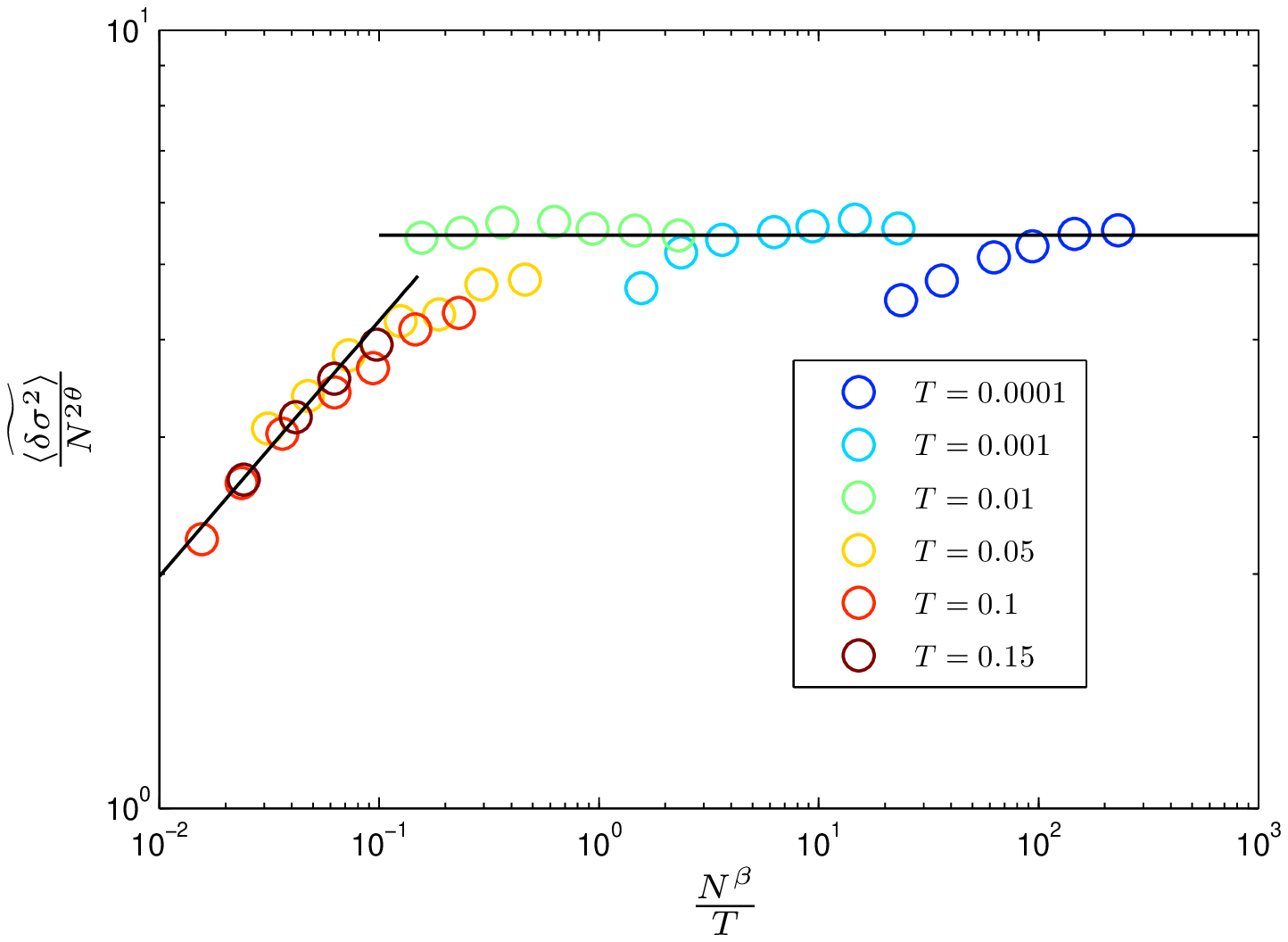}
\caption{The scaling function $g(x)$, cf. Eq. (\ref{fun}) for the 3D data (upper panel) and the 2D data (lower panel). Note the cross-over for $x$ of the order of unity as predicted by Eq. (\ref{cross2}). The power law decrease at low values of $x$ are in agreement with the prediction of $\zeta\approx 0.33$ in both cases. The two black lines represent the theoretical prediction for the scaling function $g(x)$ for $x\ll 1$ and
for $x\gg 1$.}
\label{scfunc1}
\end{figure}

To capture the temperature and size dependence of the variance, and to demonstrate unequivocally the thermal cross-over, we first need to separate the thermal from the mechanical contributions to $\langle \delta \sigma^2\rangle$. We write
\begin{equation}
\langle\delta\sigma^2\rangle = \langle\delta\sigma^2\rangle_T +
\widetilde{\langle\delta\sigma^2\rangle}\ ,
\end{equation}
where $ \langle\delta\sigma^2\rangle_T$ denotes the thermal contribution which can be read from
Eq. (10) of Ref. \cite{07IMPS}, i.e.
\begin{equation}
 \langle\delta\sigma^2\rangle_T \approx \mu T/V \ .
\end{equation}
For the mechanical part we introduce a scaling function which exhibits a cross-over
according to Eq. (\ref{cross2}). In other words, we propose a scaling function $g(x)$ to describe
the system-size and temperature dependence of the mechanical part of the variance:
\begin{equation}
\widetilde{\langle\delta\sigma^2\rangle}(N,T) =  s^2N^{2\theta} g(\bar\epsilon N^\beta/k_B T) \ .
\end{equation}
The dimensionless scaling function $g(x)$ must satisfy
\begin{eqnarray}
g(x)&\to& g_\infty;  ~{\rm for}~x\to \infty \ , \nonumber\\
g(x)&\to& g_0 x^\zeta ~{\rm for  }~ x\to 0 \ . \label{fun}
\end{eqnarray}
The first of these requirements guarantees that the fluctuation are in accordance with the athermal limit. The
second requirements guarantees that after the cross-over the fluctuations of the stress become intensive,
requiring $\zeta=-(1+2\theta)/\beta$. We compute $\zeta\approx 0.33$ both in 2D and 3D.

We present tests of the scaling function for both our 2D and 3D simulations in  Fig. \ref{scfunc1}.
Examining the scaling functions in Figs \ref{scfunc1} we see that although the data collapse
is not perfect, the thermal cross-over is demonstrated very well where expected, i.e. at values of $x$ of the order of unity. The asymptotic behavior of the scaling functions agrees satisfactorily with the theoretical prediction for both the 2D and the 3D data.

We thus conclude this letter by reiterating that the thermal cross-over
appears much more aggressive than the shear-rate cross-over in cutting off the sub-extensive scaling of the
shear fluctuations and mean drops. For macroscopic systems it should be quite impossible to observe  plastic events
that are correlated over the system size except for extremely low temperatures in the nano-Kelvin range. On the other
hand nano particles of amorphous solids may show at low temperatures and low strain rates some rather spectacular
correlated plastic events.

\acknowledgments
 This work has been supported in part by the German Israeli Foundation, the Israel Science Foundation and the Minerva Foundation, Munich, Germany. Useful discussions with Jim Langer and Eran Bouchbinder are acknowledged.


\begin{thebibliography}{99}


\bibitem{79AK}
A.S. Argon and H.Y. Kuo, Mater. Sci. Eng. {\bf 39}, 101 (1979).

\bibitem{79Arg}
A.S. Argon, Acta Metall. {\bf 27}, 47 (1979).

\bibitem{82AS}
A.S. Argon and L. T. Shi, Philos. Mag. A {\bf 46}, 275 (1982).

\bibitem{98FL}
M.L. Falk and J.S. Langer, Phys. Rev. E {\bf 57}, 7192 (1998).

\bibitem{98Sol}
P. Sollich, Phys. Rev. E, {\bf 58}, 738 (1998).

\bibitem{07BLP}
E. Bouchbinder, J.S. langer and I. Procaccia, Phys. Rev. E, {\bf 75}, 036107 (2007); {\bf 75}, 036108 (2007).

\bibitem{04ML}
C. E. Maloney and A. Lemaitre,  Phys. Rev. Lett. {\bf 93}, 016001 (2004).

\bibitem{06TLB}
A. Tanguya, F. Leonforte, and J.-L. Barrat, Eur. Phys. J. E {\bf 20}, 355 (2006).

\bibitem{06ML}
C.E. Maloney and A. Lema$\^{i}$tre, Phys. Rev. E {\bf 74}, 016118 (2006).

\bibitem{09MR}
C. E. Maloney1 and M. O. Robbins
Phys. Rev. Lett. {\bf 102}, 225502 (2009).

\bibitem{09LP}
E. Lerner and I. Procaccia, Phys. Rev. E {\bf 79}, 066109 (2009).

\bibitem{09LC}
A. Lemaitre and C. Caroli ArXiv cond-mat 0903.3196.

\bibitem{07BSLJ}
N. P. Bailey, J. Schi\o tz, A. Lema$\^{i}$tre and K. W. Jacobsen, Phys. Rev. Lett. {\bf 98}, 095501 (2007).

\bibitem{foot}
Increasing $\dot\gamma$ beyond this value causes competition with the elastic relaxation time and a disruption
of the linear shear profile.

\bibitem{07IMPS}
V. Ilyin, N. Makedonska, I. Procaccia and N. Schupper, Phys. Rev. E {\bf 76}, 052401 (2007).



\end{thebibliography}
\end{document}